\documentstyle[12pt,epsf]{article}
\newcommand{\ra}{\rangle}

\newcommand{\eps}{\epsilon}

\newcommand{\II}{{\cal I}}

\newcommand{\wt}{\widetilde}

\newcommand{\bd}{{\bar {\rm D}}}

\newcommand{\be}{\begin{equation}}
\newcommand{\ee}{\end{equation}}
\newcommand{\ben}{\begin{eqnarray}\displaystyle}
\newcommand{\een}{\end{eqnarray}}
\newcommand{\refb}[1]{(\ref{#1})}
\newcommand{\p}{\partial}
\newcommand{\sectiono}[1]{\section{#1}\setcounter{equation}{0}}

\begin{document}

{}~ \hfill\vbox{\hbox{hep-th/9906109}\hbox{MRI-PHY/P990515}
}\break

\vskip 3.5cm

\centerline{\large \bf `Blowing up' D-branes on Non-supersymmetric Cycles}  
\medskip

\vspace*{6.0ex}

\centerline{\large \rm Jaydeep Majumder and Ashoke Sen
\footnote{E-mail: joydeep@mri.ernet.in, asen@thwgs.cern.ch,
sen@mri.ernet.in}}

\vspace*{1.5ex}

\centerline{\large \it Mehta Research Institute of Mathematics}
 \centerline{\large \it and Mathematical Physics}

\centerline{\large \it  Chhatnag Road, Jhoosi,
Allahabad 211019, INDIA}

\centerline{\it and}

\centerline{\large \it International Center for Theoretical Physics}
\centerline{\large \it P.O. Box 586, Trieste, I-34100, Italy}

\vspace*{4.5ex}

\centerline {\bf Abstract}

In the orbifold limit of K3, one can give exact conformal 
field theory description
of D-branes wrapped on certain non-supersymmetric 
cycles of K3.  We study the effect
of switching on the `non-geometric blow up modes' 
corresponding to anti-symmetric
tensor gauge field flux through the 2-cycles 
on these D-branes. Working to first
order in the blow up parameter, we determine 
the region of the moduli space in which
these D-branes are stable. Across the boundary 
of this region, the D-brane wrapped
on the non-supersymmetric cycle decays to a pair 
of D-branes, each wrapped on a
supersymmetric cycle, via a second order 
phase transition. 

\vfill \eject

\tableofcontents

\baselineskip=18pt

\sectiono{Introduction and Summary} \label{ss1}

BPS D-branes\cite{9510017,DBRANE} have proved to be extremely useful in
studying
various aspects of string dualities, stringy black holes and other
properties of string theory. The original formulation of D-branes was
given for flat D-branes in flat space-time. Significant progress was made
in \cite{9603167} in the study of curved D-branes. In particular
\cite{9603167} showed how to find an exact boundary conformal field theory
description of D-branes wrapped on certain 2-cycles of K3 in the orbifold
limit. The 2-cycles studied there correspond to the cycles associated
with the blow up of the orbifold fixed points.\footnote{Throughout this
paper we shall use the words fixed points and orbifold planes
interchangeably, both refering to five dimensional
fixed planes spanning the non-compact directions.} Although in the
orbifold limit these 2-cycles have zero size, D-branes wrapped on these
2-cycles have finite tension due to the presence of the anti-symmetric two
form flux through these cycles\cite{9507012}. These cycles are
supersymmetric, so that D-branes wrapped on these cycles correspond to BPS
D-branes.


If we take two such branes, associated with two different 2-cycles ({\it
i.e.} different fixed points $P$ and $Q$ of the orbifold) then the
combined system could be
non-supersymmetric although the individual branes are BPS. It was shown in
\cite{9812031,9901014} that in certain region of the moduli space, there
is a single non-BPS D-brane configuration carrying the same charge quantum
numbers as this combined system,
but with tension less than the sum of the tension of the two individual
D-branes. Thus we can regard this single brane as a (classical) bound
state of the two BPS branes\cite{9805019,9808141,9810188}. It can also be
interpreted as a D-brane
wrapped around a single 2-cycle of K3 which is homologically equivalent to
the sum of the two individual cycles associated with the fixed points $P$ 
and $Q$. Since the wrapped D-brane is
non-BPS, the associated cycle is non-supersymmetric.
Refs.\cite{9812031,9901014}
gave an exact
boundary conformal field theory description of this non-BPS brane.

In the orbifold limit the relevant modulus which controls the stability
of the non-BPS brane is the radius $\wt R$ of the circle of the original
torus 
passing through the
orbifold fixed points $P$ and $Q$.\footnote{For simplicity we are assuming
that the torus used in the construction of the orbifold is a product of
four circles and the points $P$ and $Q$ lie
along one of the circles. We also take the radii of the other circles to
be sufficiently large in order to avoid other kinds of
instability\cite{9812031,9901014} than the ones which will be discussed
here.} When this radius is less than
a critical radius $\wt R_c$, the brane wrapped on the non-supersymmetric
cycle has
lower tension than the sum of the tensions of the two supersymmetric
branes. This is reflected in the fact that in this region of moduli
space the non-BPS brane has no
tachyonic mode and hence is stable, whereas the system containing the pair
of BPS branes has a tachyonic mode and hence is unstable. When the radius
is larger than the critical radius the situation is reversed. Now the
tension of the non-BPS brane is larger than the sum of the tension of the
two BPS branes. Furthermore the tachyonic mode disappears from the system
containing pair of BPS branes, and a tachyonic mode appears on the non-BPS
brane. Thus in this region of the moduli space the stable system is
clearly the configuration of two BPS branes.\footnote{Throughout this
paper we shall restrict our analysis to open string tree level. Thus the
process of formation of the bound state via tachyon condensation
discussed here is distinct from the bound state formation via possible
attractive force due to closed string exchange interaction. The former is 
an open string tree level effect whereas the latter is open string one
loop effect.}

At the critical radius $\wt R_c$ both systems have a massless open string
mode
representing the limit of the tachyonic mode from their respective region 
of instability. We shall refer to this as the tachyonic mode although at
the critical radius it is not tachyonic. 
One can show that this represents an exactly marginal deformation which
interpolates between the system containing a pair of BPS branes and the
non-BPS brane\cite{9808141,9812031,9903123,9904207,9905006}. Let us denote
by $\alpha$ the
parameter labelling the
marginal deformation, normalized 
so that $\alpha=0$ (mod 2) represents the pair of
BPS branes and $\alpha=1$ (mod 2) represents the non-BPS brane.
$\alpha$ can be interpreted as the vacuum expectation value (vev) of the
tachyonic mode on the brane. Away from the critical radius the tachyonic mode
develops a potential.
This potential energy
(density) $V(\alpha)$ is periodic 
under $\alpha\to\alpha+2$ due to a periodicity in
the underlying conformal field theory\cite{9808141}, and has the following
qualitative behaviour:
\begin{figure}[!ht]
\begin{center}
\epsfbox{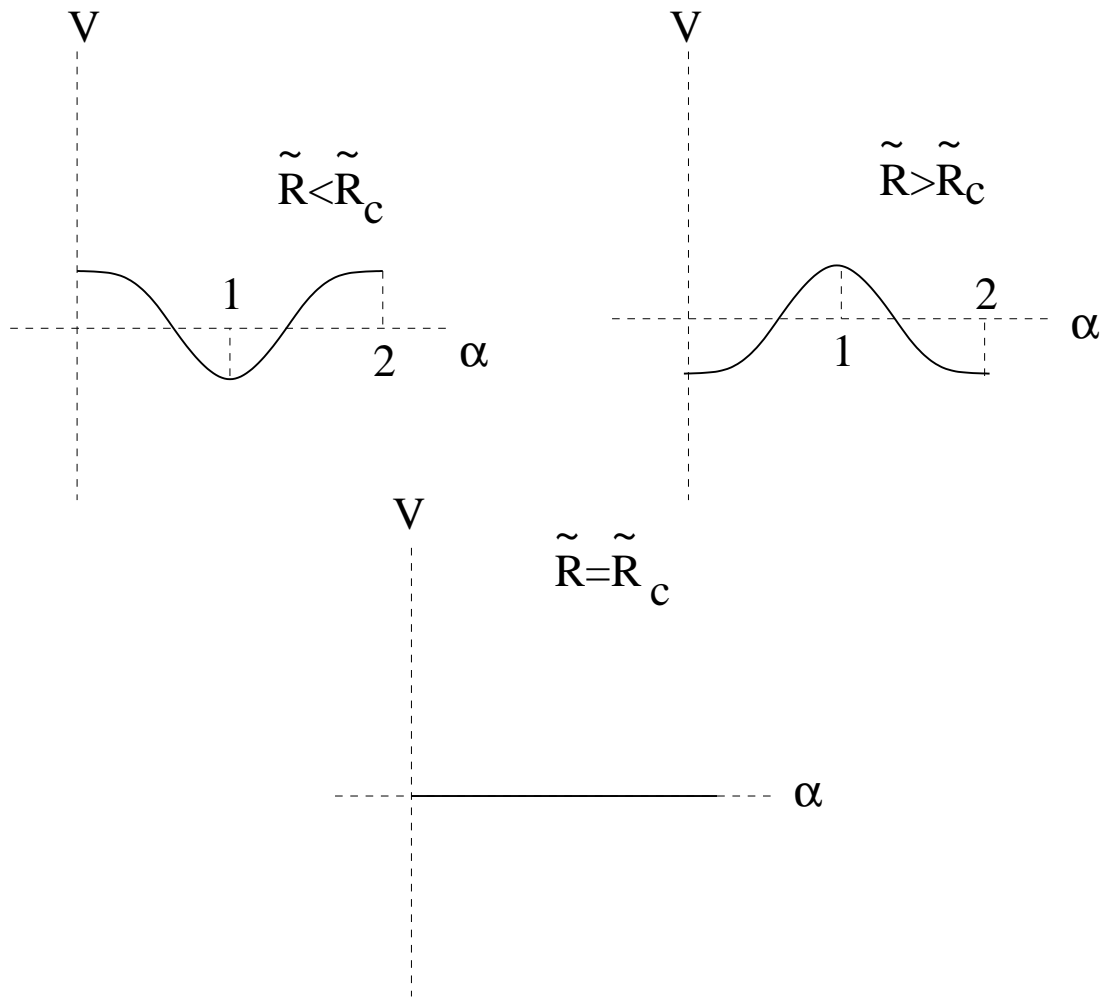}
\end{center}
\caption{The potential $V(\alpha)$ for different values of $\wt R$.}
\label{f2}
\end{figure}
\begin{figure}[!ht]
\begin{center}
\epsfbox{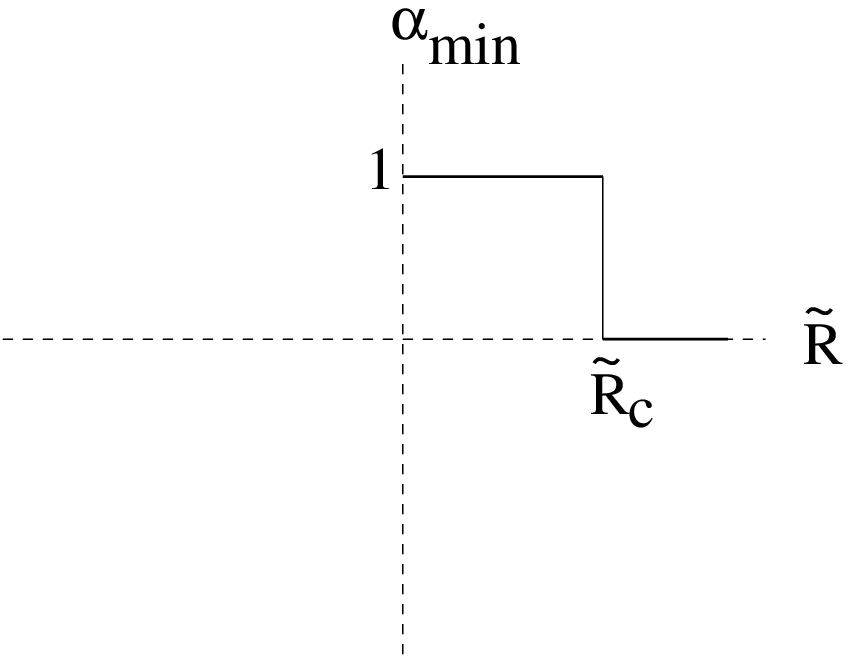}
\end{center}
\caption{Location of the minimum of $V(\alpha)$ for different values of
$\wt R$.}
\label{f4}
\end{figure}
\begin{enumerate}

\item At the critical radius $V(\alpha)$ vanishes, since the tachyonic
deformation is exactly marginal.

\item For $\wt R>\wt R_c$, $V(\alpha)$ has a minimum at $\alpha=0$
and a maximum at $\alpha=1$. This shows that the $\alpha=0$ configuration,
representing a pair of BPS branes, is the stable configuration, whereas
the $\alpha=1$ configuration representing the non-BPS brane is unstable.

\item For $\wt R < \wt R_c$, $V(\alpha)$ has a minimum at $\alpha=1$
and a maximum at $\alpha=0$, showing that the $\alpha=1$ configuration,
representing a non-BPS brane, is the stable configuration. The $\alpha=0$
configuration, being the maximum of $V(\alpha)$, is unstable.
\end{enumerate}
This has been sketched in Fig.\ref{f2}.
Note that the location of the ground state in the $\alpha$ space jumps
discontinuously as we pass through the critical radius, as shown in
Fig.\ref{f4}.
\begin{figure}[!ht]
\begin{center}
\epsfbox{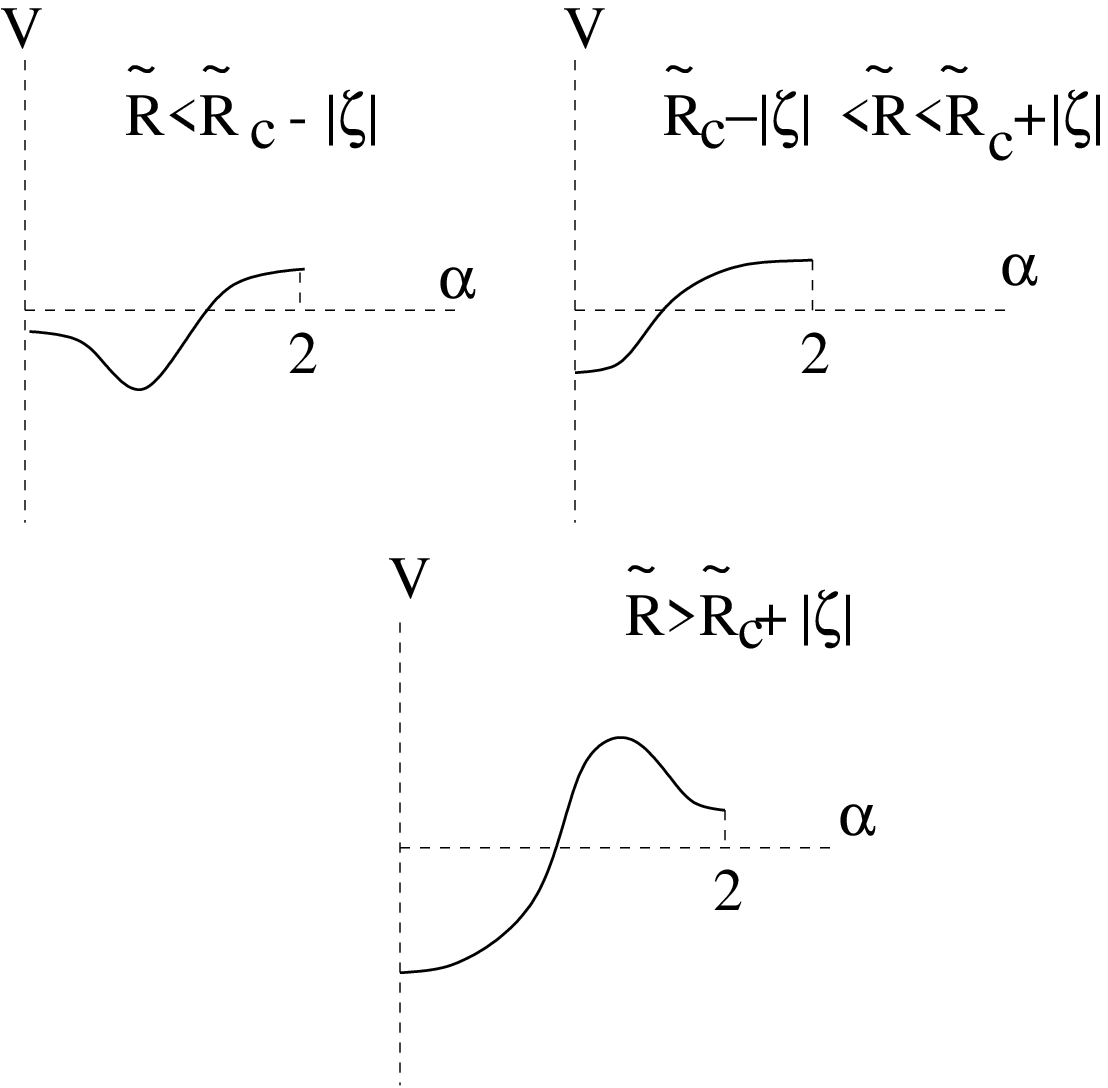}
\end{center}
\caption{The potential $V(\alpha)$ for different values of $\wt R$ for
$\zeta<0$.}
\label{f3}
\end{figure}

This is the result in the orbifold limit of the theory. In this paper we
shall study how this picture gets modified when we blow up the orbifold
fixed points. In fact the particular moduli which we shall turn on are not
the geometric blow up modes, but deformations corresponding to changing
the flux of the anti-symmetric tensor field through the
2-cycle. Although these particular moduli do not correspond to
the geometric blow up parameters, but deformations of the Kahler class
associated with the cycles by an imaginary part, we shall refer to these
as
the blow up modes. We find
that to first order in the blow up parameters, the potential $V(\alpha)$
depends on only one of  these parameters, which is the difference in the
antisymmetric tensor field flux through the two 2-cycles. We compute the
complete potential $V(\alpha)$ to first order in the blow up parameter
$\zeta$ and
first order
in the difference $(\wt R-\wt R_c)$. The
result is: 
\be \label{eintone}
V(\alpha) \propto ({1\over 4}(\wt R_c-\wt R)\cos(\alpha\pi) +
\zeta
\cos({1\over 2} \alpha \pi))\, .
\ee
{}From this we can study the
locations of the extrema of the potential for various ranges of $\wt R$.
We
can also identify the nature of the absolute minimum of $V(\alpha)$ in
different ranges of $\wt R$ by continuously connecting it to a minimum of
$V(\alpha)$ for $\zeta=0$, where the identification is known. 
As is clear from eq.\refb{eintone}, $V(\alpha)$ is invariant under 
$\alpha\to\alpha+4$, and 
$\alpha\to -\alpha$. Using this we can restrict $\alpha$ to the range
$0\le\alpha\le
2$. In this range the minimum of $V(\alpha)$ has the following structure:
\begin{figure}[!ht]
\begin{center}
\epsfbox{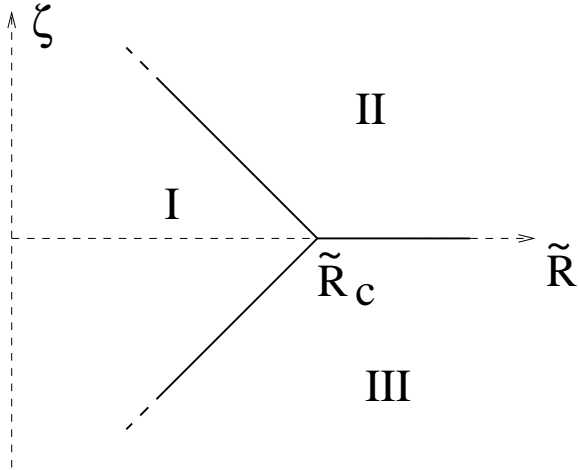}
\end{center}
\caption{Phase diagram in the $\wt R-\zeta$ plane. Phase I
($\alpha_{min}=(2/\pi)\cos^{-1}
(\zeta/(\wt R-\wt R_c))$ corresponds to a D2-brane
wrapped on a non-supersymmetric cycle, 
phase II ($\alpha_{min}=2$) corresponds to a 
pair of D2-branes, each wrapped on a supersymmetric cycle, and phase III
($\alpha_{min}=0$) corresponds to a pair of D2-branes wrapped on the same
supersymmetric cycles as in phase II, but carrying opposite D0-brane charges
compared to those in
phase II. The shape of the curves 
displayed here is valid only to first order in
$\zeta$ and $(\wt R-\wt R_c)$.} 
\label{f5} 
\end{figure}
\begin{enumerate}
\item For $\wt R > (\wt R_c - |\zeta|)$ the absolute minimum of
the
potential corresponds to a pair of BPS D-branes. This minimum is at
$\alpha=0$ ($\alpha=2$) for $\zeta<0$ ($\zeta>0$). The $\alpha=2$ configuration
differs from the $\alpha=0$ 
configuration in that the D0-brane charge carried by the
pair of wrapped membranes get exchanged.

\item For $\wt R < (\wt R_c - |\zeta|)$ the absolute minimum of
the
potential corresponds to a single non-BPS D-brane. This minimum is at
$\alpha={2\over \pi}\cos^{-1}(\zeta/(\wt R-\wt R_c))$. For $|\zeta|<<|\wt
R-\wt R_c|$ the minimum approaches $\alpha=1$ in agreement with the
answer for $\zeta=0$.

\end{enumerate}
Fig.\ref{f3} gives a sketch of the potential for various ranges of values
of $\wt R$ for $\zeta<0$. Fig.\ref{f5} shows the phase diagram in the $(\wt
R,\zeta)$ plane.

\begin{figure}[!ht]
\begin{center}
\epsfbox{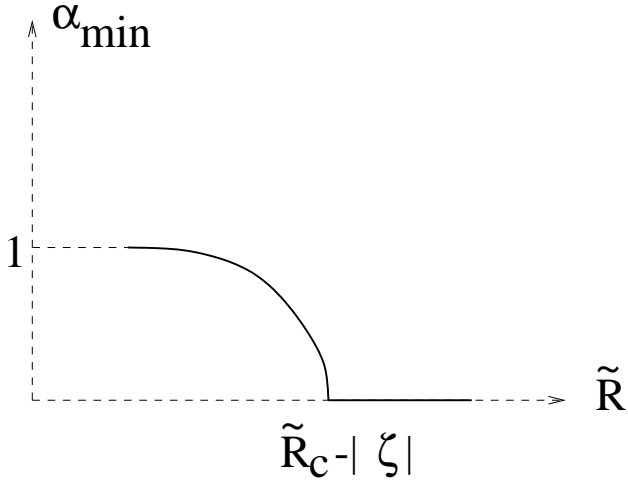}
\end{center}
\caption{Location of the minimum $\alpha_{min}$ of $V(\alpha)$ for
different values of $\wt R$
for $\zeta<0$. As $|\zeta|\to 0$, this approaches a step function.}
\label{f1}
\end{figure}
{}From this we see that for 
$\zeta\ne 0$ the critical radius is shifted to $(\wt R_c
-|\zeta|)$. Also as $\wt R$ approaches $(\wt R_c
-|\zeta|)$ from below (region I in Fig.\ref{f5}), the location of the absolute
minimum
approaches the point $\alpha=0$ ($\alpha=2$) for $\zeta<0$ ($\zeta>0$).
This is the same as the location of the minimum for $\wt R>(\wt R_c
-|\zeta|)$. 
Fig.\ref{f1} shows a sketch of
the evolution
of the minimum of $V(\alpha)$ as a function of $\wt R$ for a fixed $\zeta<0$. 
Thus {\it there is no discontinuous jump in the
location of the minimum as we pass through the phase boundary between regions I
and III and between regions I and II}. 
A detailed
analysis of the potential shows that in this case the phase transition
from the non-BPS D-brane to the pair of BPS D-branes is second
order.\footnote{Possibility of the existence of such critical points in
non-BPS D-branes was speculated by C. Vafa\cite{VAFAP}.}
On the other
hand the minimum of $V(\alpha)$ 
jumps discontinuously from $\alpha=0$ to $\alpha=2$
as $\zeta$ changes sign keeping $\wt R>\wt R_c$. Thus the phase transition
between regions II and III across
the $\zeta=0, \wt R>\wt R_c$ line is first order.

The rest of the paper is organised as follows. Although our analysis is
valid for any even (odd) dimensional D-brane in type IIA (IIB) string
theory wrapped
on (non)-supersymmetric cycles of K3, for convenience we shall focus on a
particular case $-$ D2-brane of type IIA string theory wrapped on the
cycles of K3. In section \ref{ss2} we give a description of these states
using the (non-BPS) D-branes of type II string theory, and give a precise
statement of the problem that we want to solve. In section \ref{ss3} we
solve the problem by finding the tachyon potential, and finding its
extrema.

\sectiono{Statement of the Problem} \label{ss2}

In this section we review some of the earlier results which will be
required for our analysis, carry out some preliminary analysis of the
problem, and give a precise statement of the
problem that we shall solve in the next section.
The system that we shall analyse is the same one as in \cite{9812031},
namely a
non-BPS D-string of IIA wrapped on a circle $\wt S^1$ of radius $\wt R$,
modded out by $\II_4$ where
$\II_4$ reverses the direction tangential to $\wt S^1$ and three other
directions.\footnote{The D-string is taken to be at the origin of these
other three coordinates, so that the original configuration is invariant
under $\II_4$.} This can be regarded as a two brane wrapped on a
non-supersymmetric cycle of $K3$ in the orbifold limit if the other three
directions are
compact\cite{9812031,9901014}. We shall take the radii of these three
directions
to be
sufficiently
large so that there are no tachyonic modes from open strings wound in
these directions. In order to use some of the already known results, we
shall first make a T-duality
transformation on $\wt S^1$, so that the background now represents type
IIB
string theory on the dual $S^1$ modded out by
$(-1)^{F_L}\cdot\II_4$, and
the non-BPS D-string of the original type IIA theory becomes a non-BPS
D-particle of IIB\cite{9809111,9806155} 
stuck to one of the orbifold planes. This
system is identical 
to the one analysed in \cite{9805019,9806155}, and can also be
identified to the system analysed in \cite{9808141} before the
orbifold projection. We shall denote by $R=\alpha' \wt R^{-1}$ the
radius of the dual $S^1$ and by $x$ the coordinate of the dual $S^1$ and
take the non-BPS D-particle to be located at $x=0$. We shall refer to this
new description of the system as the IIB description, and the original
description as the IIA description. {}From now on we shall continue to use
the type IIB description unless mentioned otherwise.

Although the non-BPS D-particle in type IIB string theory has a tachyonic
mode, it is projected out by the orbifolding
operation\cite{9806155,9808141}. However, as we
reduce the radius $R$ of $S^1$, the open string stretched between the
D-particle and its image across $S^1$ develops a tachyonic mode.
Let $R_c$ denote the critical radius below which
this tachyonic
mode appears. In the $\alpha'=1$ unit that we shall be using, $R_c={1\over
\sqrt 2}$.
The physical interpretation of the appearance of the tachyonic mode can be
understood by studying the mass of the non-BPS
D-particle, as well as the total mass of a D-string $\bd$-string pair of
type IIB string theory, wrapped on $S^1$.
If $g$ denotes the type IIB string coupling constant, then the mass of the
non-BPS D-particle stuck to the orbifold plane is given by:
\be \label{exthree} 
m_{D0}= {1\over 2} \cdot \sqrt 2 \cdot {1\over g} =
{1 \over \sqrt 2 g} \, . 
\ee
The various factors in this expression can be understood as follows.
Since
taking orbifold by
$\II_4\cdot (-1)^{F_L}$ cuts space into half its original size, it keeps
only half of the D-particle at $x=0$. Thus its mass is half of that of the
non-BPS D-particle in type IIB string theory, which in turn is equal to
$(\sqrt 2/g)$. By the same argument, after the orbifold projection the
mass of a D-string wrapped on $S^1$ is computed by multiplying its tension
by $\pi R$ instead of $2\pi R$, since one fundamental region of the
orbifold contains a piece of the D-string stretched from the fixed point
at $x=0$ to the fixed point at $x=\pi R$. The total
mass of the
D-string $\bd$-string pair stretched from $x=0$ to $x=\pi R$ is given by:
\be \label{exfour}
m_{D1\bd 1}= 2.{1\over 2\pi g}.\pi R = {R\over g}\, .
\ee
We see that at $R=R_c={1\over \sqrt 2}$ the two systems are degenerate.
Below the critical radius the D-string $\bd$-string
pair has lower energy. Thus it would be natural to associate the
tachyonic instability of the D0-brane to the possibility of its decay
into a D-string $\bd$-string pair, provided the two systems carry the
same charge quantum numbers. 

In order to investigate whether the D-particle carries the same charge
quantum numbers as a D-string $\bd$-string pair, we use the boundary
state\cite{POLCAI,CLNY,ONOISH,ISHIBASHI,BOUNDARYOLD,BOUNDARY}  
description of the two systems. The boundary state describing the
D-particle was constructed in \cite{9806155}. It is a linear combination
of untwisted sector Neveu-Schwarz-Neveu-Schwarz (NSNS) states and twisted
sector Ramond-Ramond (RR)
states, from which it
follows that it is charged under the gauge field at $x=0$ originating in
the
twisted RR sector. The situation with a D-string ($\bd$-string) is
somewhat more complex.
The boundary state describing a BPS D-string ($\bd$-string) is
characterized by a $Z_2$ Wilson 
line $e^{i\theta}=\pm 1$ along $S^1$, and two more
parameters
$\epsilon_1$, $\epsilon_2$ which can take values $\pm 1$\cite{9805019}:
\ben \label{exsix}
|\theta,\eps_1,\eps_2\ra &=& {1\over 2} (|\theta,U\ra_{NSNS} +\eps_1
|\theta,U\ra_{RR}) + {1\over 2\sqrt 2} \eps_2 (|T_1\ra_{NSNS} +\eps_1
|T_1\ra_{RR}) \nonumber \\
&& \qquad + {1\over 2\sqrt 2} e^{i\theta} \eps_2 (|T_2\ra_{NSNS} +\eps_1
|T_2\ra_{RR})\, .
\een
Here  $U$ stands for untwisted sector, $T_1$ stands
for twisted sector at $x=0$ and $T_2$ stands for twisted sector at $x=\pi
R$. $\eps_1$ takes value +1 ($-1$) for a D-string ($\bd$-string), and
$\eps_1\eps_2$ denotes the sign of the twisted sector RR charge carried by
the
$x=0$ end of the D-string. 
$e^{i\theta}\eps_1\eps_2$ denotes the sign of the twisted
sector RR charge carried by the $x=\pi R$ end of the string.

Thus we see that
after modding out by
$(-1)^{F_L}\cdot\II_4$ a
D-string ($\bd$-string) carries twisted sector Ramond-Ramond (RR) charge
at its two
ends. If we choose a D-string state that carries + charge at the $x=0$
end and $-$ charge at the $x=\pi R_c$ end corresponding to
$(\theta,\eps_1,\eps_2)=(\pi,+,+)$ (configuration (a) in Fig.6 of
\cite{9805019}),
and a $\bd$-string that  carries $+$ charge at both ends
corresponding to $(\theta,\eps_1,\eps_2)=(0,-,-)$ (configuration
(g) of Fig.6 of \cite{9805019}), then the
D-string $\bd$-string pair is neutral under the twisted sector gauge field
at $x=\pi R_c$ and carries +ve charge under the twisted sector gauge field
at $x=0$. This matches with the charge quantum number of the non-BPS
D-particle at $x=0$, as can be seen from the fact that the
massless RR component of the boundary state describing a
non-BPS D-particle at $x=0$\cite{9806155}(published
version) is given precisely by twice
that appearing in eq.\refb{exsix}.\footnote{This relative 
factor of $(1/2)$ between the twisted
sector RR charge carried by the end of a BPS D-string and by a non-BPS
D-particle can be seen as follows. The open string partition function 
for the non-BPS D-string has a projection operator
${1+(-1)^F\over 2}{1+g_1\over 2}{1+g_2\over 2}$, where $F$ denotes
world-sheet fermion number, $g_1$ denotes $(-1)^{F_L}$ accompanied by 
the transformation $(x^6.\ldots x^9\equiv x)\to (-x^6, \ldots -x^9)$, 
and $g_2$ denotes $(-1)^{F_L}$ accompanied by 
the transformation $(x^6.\ldots x^9)\to (-x^6, \ldots -x^8, 
2\pi R-x^9)$\cite{9805019}. On the other hand, partition function of open
strings living on a
non-BPS D-particle at $x\equiv x^9=0$ has a projection operator
${1+(-1)^F\cdot g_1\over 2}$\cite{9806155}. Thus the coefficient of
$(-1)^F\cdot g_1$ in
the two cases differ by a factor of 4. Since this coefficient is related
to the norm of the twisted sector RR component of the boundary state
describing the system, we conclude that the twisted sector RR component of
the boundary state describing a D-particle has an extra factor of 2
compared to that describing a D-string.} Hence the D-particle can decay
into this pair of D-string states. The appearance of the tachyonic mode
on the D-particle below $R=R_c$ signals the possibility of this decay.
{}From this analysis we see that this configuration requires the D-string
to
carry a $Z_2$ Wilson line, whereas the $\bd$-string does not carry any
Wilson line. In the
dual type IIA description the D-string $\bd$-string pair, with the
D-string carrying a $Z_2$ Wilson line, corresponds to a D0-$\bd$0 pair
situated at diametrically opposite points on the dual circle $\wt S^1$.
After
modding out by $\II_4$ this can be reinterpreted as a pair of BPS
D2-branes of IIA,
wrapped on the two cycles associated with the two orbifold fixed points
on $\wt S^1$\cite{9603167}.

Note that we could also consider 
a D-string state that carries + charge at both ends
corresponding to
$(\theta,\eps_1,\eps_2)=(0,+,+)$ (configuration (c) in Fig.6 of
\cite{9805019}),
and a $\bd$-string that  carries $+$ charge at the $x=0$ end and 
$-$ charge at the $x=\pi R_c$ end 
corresponding to $(\theta,\eps_1,\eps_2)=(\pi,-,-)$ (configuration
(e) of Fig.6 of \cite{9805019}). This has the same RR charge as the
previous configuration, and so the D-particle can also decay into this
state. This differs from the previous configuration by having the Wilson
line on the $\bd$-string rather than on the D-string.  In the dual type
IIA description this again corresponds to a D0-$\bd$0 pair on $\wt
S^1$, but with their positions reversed.

A more systematic analysis of the transition from the D-particle state to
a  D-string $\bd$-string pair was carried out in
refs.\cite{9808141,9812031}. It was shown that
the lowest mode of the tachyon, which is massless at
$R=R_c$, represents an exactly marginal deformation. By switching on this
marginal deformation one can continuously interpolate between the
boundary conformal field theories (BCFT) describing the
non-BPS D-particle and the D-string $\bd$-string pair. We could see this
marginal deformation either by starting from the D-particle side, or
by starting from the D-string $-$ $\bd$-string side.  We shall find it
more convenient to
do the analysis from the D-string $\bd$-string side, so that we can use
the results of \cite{9808141}. 

On an infinite D-string $\bd$-string system  there is a tachyonic mode
with mass$^2=-{1\over 2}$. 
Upon wrapping the D-string $\bd$-string pair 
on a circle of radius $R$, with a 
$Z_2$ Wilson line on
the D-string, the tachyon field $T$ is anti-periodic under $x\to x+2\pi
R$, and has a Fourier expansion of the form:
\be \label{exone}
T(x)=\sum_{n\in Z} T_{n+{1\over 2}} e^{i(n+{1\over 2}){x\over R}}\, .
\ee
The effective mass$^2$ of the mode $T_{n+{1\over 2}}$ is given by
\be \label{extwo}
m^2_{n+{1\over 2}}={(n+{1\over 2})^2\over R^2} -{1\over 2}\, .
\ee
Thus for $R\le R_c={1\over\sqrt 2}$ there are no tachyonic modes on this
system.
For $R>R_c$, $T_{\pm{1\over 2}}$ becomes tachyonic,
indicating that the system becomes unstable
in this range of $R$.
At $R=R_c$, $(T_{1\over 2}\pm T_{-{1\over 2}})$ can be shown to be exactly
marginal, and
one can deform the BCFT by switching on vev of this field. But from
eq.(2.59) of \cite{9805019} one finds that only the
mode $(T_{1\over 2}-T_{-{1\over 2}})$ is invariant under
$(-1)^{F_L}\cdot\II_4$ and can be switched on.
It was shown in \cite{9808141} that by switching on the
vev of $(T_{1\over 2}-T_{-{1\over 2}})$ we can reach the BCFT describing 
the non-BPS D-particle of type IIB string theory.\footnote{Ref.\cite{9808141}
introduced the vertex operators $V_\pm$ for $T_{\pm{1\over 2}}$, which, in
the
$-1$ picture\cite{FMS}, were taken to be proportional to $\pm e^{(\pm
i/\sqrt 2)X}$.
But
if
$T_r$ is to label the $r$th mode of the tachyon field $T(x)$, then its
vertex operator in the $-1$ picture should be proportional to
$e^{irX/R}$ without any extra $r$ dependent sign. For this reason
the vertex operator for $(T_{1\over 2}-T_{-{1\over 2}})$ of
this
paper corresponds to what was called $(V_++V_-)$ in
\cite{9808141}.} 
We denote by $\alpha$
the suitably normalized vev of this
mode of the tachyon at $R=R_c$, with $\alpha=0$ representing the D-string
$\bd$-string system, and $\alpha=1$ representing the D-particle, as in
\cite{9808141}.

The marginality of $(T_{1\over 2}-T_{-{1\over 2}})$ can be seen by
rewriting
the BCFT at $R=R_c$ in terms of a different set of variables.
As discussed in \cite{9808141}, the effect of switching on the tachyon vev
is to insert the following operator at the boundary of the world-sheet:
\be \label{esix}
Tr\exp(i{\alpha\over 2\sqrt 2}\sigma_1\int dt \p_t \phi_B)\, .
\ee
Here $\int dt$ denotes integration along the boundary of the
world-sheet, $\sigma_1=\pmatrix{0 & 1\cr 1 & 0}$ is a Chan Paton
factor, $Tr$ denotes trace over the 
Chan Paton factors, and $\phi_B$ denotes the
boundary value of the world-sheet field
$\phi=\phi_L+\phi_R$,
where the field
$\phi$ is related to the bosonic coordinate field $X=(X_L+X_R)$ along
$S^1$ and its right- and left-moving world-sheet partners $\psi$, $\wt
\psi$, through the fermionization $-$ bosonization relations:
\be \label{eseven}
e^{i\sqrt 2 X_R} ={1\over \sqrt 2} (\xi + i\eta), \qquad
e^{i\sqrt 2 X_L} ={1\over \sqrt 2} (\wt\xi + i\wt\eta), 
\ee
\be \label{eeight}
e^{i\sqrt 2 \phi_R} ={1\over \sqrt 2} (\xi + i\psi), \qquad
e^{i\sqrt 2 \phi_L} ={1\over \sqrt 2} (\wt\xi + i\wt\psi).
\ee
$\xi$, $\eta$, $\wt\xi$, $\wt\eta$ are world-sheet fermion fields.
\refb{esix} can be interpreted as a Wilson line along the bosonic
coordinate $\phi$, and clearly represents a marginal deformation.

Since at $R=R_c$ the tachyon becomes marginal, the tachyon potential
$V(\alpha)$ vanishes at $R=R_c$. For $R<R_c$, $V(\alpha)$ has a minimum at
$\alpha=0$, representing the fact that the D-string $\bd$-string
configuration corresponds to the minimum energy configuration, whereas for
$R>R_c$, $V(\alpha)$ has a minimum at $\alpha=1$ indicating that the
non-BPS D-particle represents the minimum energy configuration. Thus the
point $R=R_c$ marks the phase boundary between the stable D-particle
configuration, and the stable D-string $\bd$-string configuration. 
The question that we shall be interested in is: how does this picture get
modified when we go to a different point in the moduli space of K3 (in the
original type IIA description)? We shall only analyse the effect of small
deformations
around the original configuration. These may be divided into two classes:
moduli of type IIA string theory on the torus (constant metric and
antisymmetric tensor field background), and the blow up modes of the fixed
points corresponding to twisted sector closed string states from the
NSNS sector.  There are four such blow up modes from each orbifold fixed
point. Three of these modes correspond to geometric blow up parameters,
and the fourth one corresponds to antisymmetric tensor field flux through
the 2-cycle associated with the fixed point. In the dual type IIB
description that we have been using, we have the moduli
corresponding to constant metric and anti-symmetric tensor field
background in the dual torus, and the twisted sector modes. In this case
however the twisted sector modes from the NSNS sector have a different
interpretation. Each orbifold plane obtained by modding out by
$\II_4\cdot(-1)^{F_L}$ has a hidden NS five brane, since it is
S-dual to the coincident orientifold 5-plane D-5-brane
system\cite{9604070}. Switching
on the twisted sector modes associated with a given orbifold plane
corresponds to moving the NS five-brane away from the orbifold plane.

It can be easily verified that switching on the constant metric or
anti-symmetric tensor field background does not modify the physics at the
phase boundary between a stable configuration of D-particle and the
D-string $\bd$-string system. To see this note that if we denote by $R$
the radius of $S^1$ measured in the new metric, then in terms of $R$, the
tachyon mass formula \refb{extwo} as well as the D-brane mass formulae
\refb{exthree}, \refb{exfour} remain unchanged. Thus at the critical
radius $R={1\over \sqrt 2}$ the D-particle becomes degenerate with the
D-string $\bd$-string system, and the tachyonic modes $T_{\pm{1\over 2}}$
become massless. We can use the same bosonization and fermionization
formulae to show that at $R=R_c$, $(T_{1\over 2}-T_{-{1\over 2}})$
represents an exactly marginal deformation and interpolates between the
BCFT describing these two systems.

Thus it remains to study what happens when we switch on the twisted sector
massless NSNS fields. We shall now argue that only
one of the eight blow up modes associated with the two fixed points 
actually affect
the masses of the D-brane system to first order and hence could modify the
physics at the phase boundary to this order. To do this we use the
boundary state description of
the D-string $\bd$-string system given in \cite{9805019} and that of the
non-BPS D-particle given in \cite{9806155}. First of all, the boundary
state
describing the non-BPS D-particle is a linear combination of the boundary
state
from untwisted sector NSNS sector, and twisted sector RR
sector\cite{9806155}. Since it
has no component from the twisted sector NSNS sector, we see that the
D-particle does not have a direct coupling to these states and hence its
mass does not depend on these twisted sector moduli to first order. 
On the other hand, from eq.\refb{exsix} we see that the D-string
($\bd$-string) 
boundary state has components along the twisted sector NSNS states, and so
the mass of the D-string ($\bd$-string) has linear dependence on the
particular moduli fields which appear in the boundary state. There are two
such moduli, one from $|T_1\ra_{NSNS}$ and the other from
$|T_2\ra_{NSNS}$. The physical
interpretation is quite clear. These moduli represent the motion of
the
NS five branes along the circle $S^1$, and since the ends of the D-string
($\bd$-string) lie on the NS five-branes, their lengths and hence their 
masses depend on the location
of the five-brane along $S^1$.\footnote{Moving the five-branes in
directions
transverse to $S^1$ does not affect the length and hence the mass of the
D-string ($\bd$-string) to first order.} 
Also note that for a given change in the location of the five brane the
change in the mass can have either sign (as reflected in the coefficient
of $|T_1\ra_{NSNS}$ and $|T_2\ra_{NSNS}$ in \refb{exsix}, which can be of
either sign). This can be traced to the fact that the D-string
($\bd$-string) can either end on the NS 5-brane or its image under
$\II_4\cdot (-1)^{F_L}$.\footnote{The S-dual version of this was explained
in Fig.2 of \cite{9805019}.} As we move the five brane in one direction,
its
image moves in the opposite direction. Thus whether the mass increases or
decreases for a given movement of the five-brane is determined by whether
the D-string
($\bd$-string) ends on the five-brane or its image.

As has been discussed earlier, the D-string $\bd$-string system under
consideration corresponds to the configurations (a) and (g) of
ref.\cite{9805019}, 
characterized by $(\theta,\eps_1,\eps_2)$ values $(\pi,+,+)$
and $(0,-,-)$ respectively. Using eq.\refb{exsix} we see that the twisted
sector component of the boundary state of the combined system is given by:
\be \label{efour}
{1\over \sqrt 2} (|T_1\ra_{RR} - |T_2\ra_{NSNS})\, .
\ee
The component $|T_1\ra_{RR}$ indicates that it carries
twisted sector RR charge associated with the $x=0$ plane, as must be the
case since it has the same charge as the D-particle situated at $x=0$. On
the other hand, the component $|T_2\ra_{NSNS}$ indicates that it couples
to NSNS sector twisted sector modes associated
with the
orbifold plane at $x=\pi R$. Thus the mass of the combined system
depends only on the location of the NS five-brane associated with the
orbifold plane at $x=\pi R$. This can be explained  by taking both the
D-string and the $\bd$-string to end on the NS 5-brane at $x=\pi R$, but
having one of them end on the 5-brane and the other end on its image near
$x=0$.\footnote{This must be the case if both of them have to carry the
same charge at the $x=0$ end and opposite charge at the $x=\pi R$ end,
since the D-string and the $\bd$-string, ending on the same five-brane,
carry opposite charge. On the other hand if the D-string and the
$\bd$-string end on the five brane and its image respectively, then their
ends carry the same charge, since the gauge field on the
five-brane world-volume that
survives the orbifold projection is the difference between the U(1) gauge 
field on the five brane and its image. This is exactly analogous to the
situation describing a D5-brane near an orientifold 5-plane.} In that case
the motion of the NS 5-brane near $x=0$ will not
change the total mass of the system, but the motion of the NS 5-brane near
$x=\pi R$ will change the total mass.

Let us denote by $\zeta$ the
specific
massless mode that appears in $|T_2\ra_{NSNS}$. {}From eq.\refb{efour} 
we see that this is the only NSNS twisted sector mode
on which the mass of the D-string $\bd$-string pair depends to first
order,
and we shall study how the transition between the D-particle state and the
D-string $\bd$-string state is affected upon switching on this mode. In
the original type IIA description,
$\zeta$ denotes the difference in the flux of the antisymmetric tensor
field through the two 2-cycles of K3, associated with the two orbifold
fixed
points.

Our
strategy will be to determine the tachyon potential completely for
$R\simeq R_c$, $\zeta\simeq 0$ to linear order in $(R-R_c)$ and $\zeta$,
ignoring terms quadratic in $(R-R_c)$ and $\zeta$, as well as
terms of order $(R-R_c)\zeta$, and then study its minimum as a function
of $\alpha$ for various values of $R$ and $\zeta$. Thus the general form
of the potential will be
\be \label{efive}
V(\alpha)\simeq{1\over g} [(R-R_c) f(\alpha) + \zeta g(\alpha)]\, ,
\ee
where $f(\alpha)$  and
$g(\alpha)$ are two functions to be determined.
Note that since we are working close to the point $R=R_c$, $\zeta=0$,
we can continue to use
the parameter $\alpha$ to label the nearly massless tachyonic mode.
Also note that we have extracted an overall power of the inverse string
coupling
($g^{-1}$) outside the potential since the $g$ dependence of the
world-volume
action of a D-brane always comes through such an overall multiplicative
factor.

In the next section we shall determine the functions $f(\alpha)$ and
$g(\alpha)$, and 
study the extremum of the potential as a function of $\alpha$. By
studying
how the minimum of the potential varies as we change the parameters $R$
and $\zeta$, we shall be able to determine the phase diagram of the
D-brane system under study in the $R-\zeta$ plane.

\sectiono{Determination of the Tachyon Potential and the Phase Diagram}
\label{ss3}

First we shall determine $f(\alpha)$. For this we set $\zeta=0$, so that
the tachyon potential has the form:
\be \label{eone}
V(\alpha) = {1\over g}[(R-R_c) f(\alpha) + O((R-R_c)^2)]\, ,
\ee
for $R\simeq R_c$.
We can determine $f(\alpha)$
by noting that $(\p V/\p\alpha) \simeq g^{-1} (R-R_c) f'(\alpha)$
corresponds to
the one point function of the tachyon to order $(R-R_c)$. This was
computed in \cite{9808141} and the answer was found to be proportional to
$\sin\pi\alpha$. Integrating this we see that $f(\alpha)$ must be
proportional to $\cos(\pi\alpha)$. The constant of proportionality can
also
be
easily found by noting that the difference between $V(\alpha)$ at
$\alpha=0$ and at $\alpha=1$ must be equal to the difference between the
total mass of the D-string $\bd$-string pair wrapped on $S^1$, and the
mass of the non-BPS D-particle. Using eqs.\refb{exthree}, \refb{exfour} we
get
\be \label{etwo}
V(\alpha=0) - V(\alpha=1) = {1\over g}(R-R_c)\, .
\ee
This gives\footnote{This analysis determines $V(\alpha)$ up to an additive
$\alpha$-independent constant which has no relevance for finding the
extrema of $V(\alpha)$ in the $\alpha$-space.}
\be \label{ethree}
f(\alpha) = {1\over 2} \cos(\alpha\pi)\, .
\ee
{}From eqs.\refb{eone} and \refb{ethree} we see that the minimum of $V$ is
at $\alpha=0$ for $R<R_c$, and is at $\alpha=1$ for $R>R_c$. This is
consistent with the fact that the D-string $\bd$-string pair is the stable
configuration for $R<R_c$, and the D-particle is the stable configuration
for $R>R_c$.

We shall now use a shortcut for determining $f(\alpha)$, which we shall
generalise later for determining $g(\alpha)$.
In the
derivation given above, we have used the tachyon one point function to
first order in $(R-R_c)$ to
compute $f'(\alpha)$.  This represents a two point function on the disk in
the theory at $R=R_c$, with one insertion of the closed string vertex
operator $V_B$ corresponding to radius deformation at the center, and one
insertion of the
tachyon
vertex operator at the
boundary\cite{9808141}. We need to choose the picture numbers\cite{FMS} of
these vertex operators so that the
total picture
number is $-2$. Let us take $V_B$ to be in $(-1,-1)$ picture, and the
tachyon vertex operator in the 0-picture. Besides this there is an
insertion of the exponential of the integrated zero picture tachyon vertex
operator at the boundary as given in eq.\refb{esix}.  Since the $\alpha$
dependence
of the one point function comes from only the matter part of the
correlation function, let us restrict ourselves to this sector. In this
case it is clear that if we start from an amplitude where we only have the
insertion of $V_B$ at the center, and the exponential of the integrated
tachyon vertex operator at the boundary, then by differentiating it with
respect to $\alpha$ we bring down an extra factor of tachyon vertex
operator at the boundary. Since this two point function has
been argued to be proportional to $f'(\alpha)$, we
see that the original amplitude is proportional to $f(\alpha)$ itself.

The main lesson from here is that $f(\alpha)$ may be computed directly by
computing the matter part of the disk amplitude with a single insertion of
$V_B$ at the
center (reflecting that we are working to order $(R-R_c)$) and insertion
of the exponential of integrated tachyon vertex operator at the boundary
given in eq.\refb{esix}. The computation of $g(\alpha)$ will be done in
the same way, with $V_B$ replaced by the appropriate twisted sector vertex
operator $V_{TW}$.\footnote{If one could construct the boundary state
describing the system at $R=R_c$ for all $\alpha$ analogously to
ref.\cite{9903123}, then one could read out $g(\alpha)$ by
computing the component of the boundary state along $V_{TW}$.}

Let us now turn to the determination of $g(\alpha)$.
Since we have already argued
that the energy of the D-string $\bd$-string system depends linearly on
$\zeta$, we see that $g(0)$ must be a non-zero constant. We shall absorb
this constant into the definition of $\zeta$ and set
$g(0)=1$.\footnote{With 
this normalization the $\zeta$ dependent
contribution to the mass of the D-string $\bd$-string pair is given by
$(\zeta/g)$. This should be equated to the total tension $(1/\pi g)$ of
the
D-string $\bd$-string pair multiplied by the shift in the position of the
NS 5-brane. Thus $\pi\zeta$ measures the shift in the position of the NS
5-brane.\label{ff1}}
On the other hand by analysing the boundary state describing the non-BPS
D-particle we have argued before that the mass of this D-particle does not
depend on $\zeta$. Hence $g(1)$ must vanish. 

The complete determination of $g(\alpha)$ 
is done by computing the disk amplitude with an insertion of the
$\zeta$ vertex operator at the center, and the exponential of
the integrated tachyon vertex operator at the boundary. We use the
notation and the normalization conventions of \cite{9808141}. 
The vertex operator $V_{TW}$ for a twisted sector state associated with
the
orbifold
plane at $x=\pi R_c$ has the property that as we go around such a vertex
operator on the fundamental string world-sheet, the various world-sheet
fields undergo the following changes:
\be \label{enine}
X\to (2\pi R_c - X), \qquad \psi\to -\psi, \qquad \wt\psi\to -\wt\psi\, .
\ee
Using eqs.\refb{eseven}, \refb{eeight} we see that
this transformation is equivalent to,
\be \label{eten}
\xi\to -\xi, \qquad \wt\xi\to -\wt\xi, \qquad \psi\to -\psi, \qquad
\wt\psi\to -\wt\psi, \qquad \eta\to\eta, \qquad \wt\eta\to \wt\eta\, ,
\ee
or to,
\be \label{eeleven}
\phi_L \to \phi_L+{\pi\over \sqrt 2}, \qquad
\phi_R \to \phi_R +{\pi\over \sqrt 2}\, .
\ee
Thus as we go around $V_{TW}$ on the string world-sheet,
$\phi=\phi_L+\phi_R$ changes by $\sqrt 2\pi$. Since $\int dt\p_t\phi_B$
measures
the total change of $\phi$ as we go around the boundary of the disk, we
see that if there is an insertion of $V_{TW}$ at the center of the disk,
then 
\be \label{etwelve}
Tr\exp(i{\alpha\over 2\sqrt 2}\sigma_1\int dt \p_t \phi_B)
=Tr\exp(i{\alpha\over 2\sqrt 2} \cdot \sigma_1\cdot \sqrt 2 \pi) = 2
\cos({1\over 2} \alpha\pi)\, .
\ee
This shows that $g(\alpha)$ is proportional to $\cos({1\over 2}
\alpha\pi)$. Using the normalization $g(0)=1$, we get
\be \label{ethirteen}
g(\alpha) = \cos({1\over 2}\alpha\pi)\, .
\ee
This satisfies the condition $g(1)=0$ derived earlier.
Thus the full tachyon potential to this order is given by:
\be \label{efourteen}
V(\alpha) \simeq{1\over g}({1\over 2}(R-R_c)\cos(\alpha\pi) + \zeta
\cos({1\over 2} \alpha \pi))\, .
\ee

Note that the potential is periodic in $\alpha$ with periodicity 4. This
may appear as a surprise, as in ref.\cite{9808141} it was found that the
BCFT describing the D-string $\bd$-string system is periodic in $\alpha$
with periodicity 2. We can understand the origin of this apparent
discrepancy as follows. Let us set $\zeta=0$, $R=R_c$, and start from the
D-particle
state represented by the point $\alpha=1$. We can perturb this
system by the marginal tachyonic deformation and study the fate of the
BCFT as a
function of the new deformation parameter $(\alpha-1)$. The T-dual version
of this analysis in the type IIA description was carried out in
\cite{9812031}. In this analysis the
starting configuration was a non-BPS D-string of IIA wrapped on a circle.
Switching on the marginal deformation corresponding to
$(\alpha-1)=1$
corresponds to the creation of a kink-antikink pair on the circle, which
is
to be interpreted as a D0-brane $\bd$0-brane pair of type IIA string
theory situated at diametrically opposite points on a 
circle\cite{9812031,9812135}.\footnote{Note that $(\alpha-1)$ was denoted
by $\alpha$ in
\cite{9812031}, since there the starting configuration was the non-BPS
D-string.} On the other
hand if we take $(\alpha-1)=-1$, the effect is to create an antikink-kink
pair. Thus the result is again a D0-$\bd$0 pair, but with their positions
reversed. This has the following interpretation in the dual type IIB
description. If we take the $\alpha=0$ configuration to represent a
D-string
$\bd$-string pair with a Wilson line on the D-string, then the $\alpha=2$
configuration denotes a D-string $\bd$-string pair with a Wilson line
along the $\bd$-string. These correspond to the same BCFT before the
orbifold projection, but differ in the orbifold theory when twisted
sector modes are switched on. As discussed
in section \ref{ss2}, since the Wilson line is on the $\bd$ string,
the $\alpha=2$ configuration
corresponds to the pair of states carrying
$(\theta,\eps_1,\eps_2)$ quantum numbers $(0,+,+)$ and $(\pi,-,-)$
respectively  (pair of states (c) and (e) in
the language of \cite{9805019}). Using eq.\refb{exsix} we see that the
twisted sector
component of
the boundary state
describing the combined system is given 
by:
\be \label{efourxx}
{1\over \sqrt 2} (|T_1\ra_{RR} + |T_2\ra_{NSNS})\, .
\ee
Comparing with eq.\refb{efour} we see that it carries the same twisted
sector RR charge as the system at $\alpha=0$, but its coupling to the
twisted sector NSNS state is opposite to that of the system at
$\alpha=0$.
Thus the $\zeta$ dependent component of the tachyon potential
should have opposite signs at $\alpha=0$ and at $\alpha=2$, as is the case
for the potential given in eq.\refb{efourteen}. 

Using the periodicity
$\alpha\to \alpha+4$ we can restrict $\alpha$ to the range $-2< \alpha\le
2$. Also, by making a gauge transformation on the D-string $\bd$-string
system at $\alpha=0$, we can change the sign of the tachyon, which
corresponds to the transformation $\alpha\to -\alpha$. Thus $\alpha$ and
$-\alpha$ denote equivalent configurations, and the physical range of
$\alpha$ can be taken to be $0\le\alpha\le 2$.
Finally,
without any loss of generality we can take $\zeta$ to be negative in our
analysis, since the sign of $\zeta$ can be changed by a redefinition
$\alpha\to 2-\alpha$. 

Let us define:
\be \label{efifteen}
u = 2(R-R_c)\, .
\ee
By analysing the potential \refb{efourteen} with $\zeta<0$, we get the
following results:
\begin{enumerate}
\item For $u<(-|\zeta|)$, $V(\alpha)$ has a pair of minima at $\alpha=0$
and
at $\alpha=2$, and a maximum at
$\alpha=(2/\pi)\cos^{-1}(-\zeta/u)$. The absolute minimum is at
$\alpha=0$. As $\zeta\to 0$, this state goes over smoothly to the
D-string $\bd$-string system 
with the Wilson line on the D-string. Thus even for
non-zero $\zeta$ we can
identify the system at $\alpha=0$ to a D-string $\bd$-string
pair.

\item For $(-|\zeta|)<u<(|\zeta|)$, $V(\alpha)$ has a minimum at
$\alpha=0$
and a maximum at $\alpha=2$. As $u$ passes through the point
$u=-|\zeta|$, 
the minimum at $\alpha=0$ evolves smoothly. Thus even in this range of
$u$, we can interprete the stable minimum at $\alpha=0$ as a
D-string $\bd$-string pair.

\item For $u>(|\zeta|)$, $V(\alpha)$ has a pair of maxima at $\alpha=0$
and at
$\alpha=2$, and a minimum at 
$\alpha=(2/\pi)\cos^{-1}(-\zeta/u)$. As $\zeta\to 0$, this minimum
evolves smoothly to the stable non-BPS D-particle corresponding to the
point $\alpha=1$. Thus by continuity we can conclude that for $u>|\zeta|$ 
the stable
minimum at $\alpha=(2/\pi)\cos^{-1}(-\zeta/u)$ denotes the
stable non-BPS D-particle.

\end{enumerate}

The results for $\zeta>0$ can be obtained by using the symmetry
$\zeta\to -\zeta$, $\alpha\to 2-\alpha$. 
In this case the $\alpha=2$ configuration,
representing a D-string $\bd$-string pair with Wilson line on the $\bd$ string,
corresponds to the stable minimum for $u<|\zeta|$, and the $\alpha={2\over
\pi}\cos^{-1}(-{\zeta\over u})$ 
configuration, representing a non-BPS D-particle at
$x=0$, corresponds to the stable minimum for $u>|\zeta|$.

If we want to translate these results to the dual IIA description, we only
need to note that the radius $\wt R$ of the circle in this description is
given by $(1/R)$. Thus $(\wt R -\wt R_c)\simeq {1\over R_c^2}(R_c-R) =
2(R_c-R)$. Thus the parameter $u$ can be identified as $(\wt R_c-\wt
R)$. This reproduces the results quoted in the introduction. 

The phase diagram in the $u-\zeta$ plane is quite simple. For $\zeta<0$,
$u<(|\zeta|)$ the D-string $\bd$-string with Wilson line on the D-string is the
stable configuration. In
the dual theory describing type IIA on K3, this corresponds to a pair of
D2-branes, each wrapped on a supersymmetric cycle of K3. For $\zeta>0$,
$u<|\zeta|$, the D-string $\bd$-string 
system with Wilson line on the $\bd$-string 
is the stable configuration. In the dual IIA theory this again corresponds to a
pair of wrapped D2-branes, 
but carrying opposite D0-brane charges compared to the
previous configuration. 
For $u>(|\zeta|)$ the non-BPS D-particle is the stable configuration for all
$\zeta$. In the
dual type IIA theory this represents a D2-brane wrapped on the
non-supersymmetric cycle. The phase diagram in the 
$(\wt R=\wt R_c-u,\zeta)$ plane
has been shown in Fig.\ref{f5}.

The location of the minimum $\alpha_{min}$ of
$V(\alpha)$ evolves continuously as $u$ crosses the phase boundary
$|\zeta|$.
It is instructive to study the nature of the transition across the line
$u=|\zeta|$. For this note that for $\zeta<0$, $u>|\zeta|$ the potential has
two maxima and a minimum in the range $0\le\alpha\le 2$. As $u$ approaches
$|\zeta|$ from above, the maximum at $\alpha=0$, the minimum
at $\alpha=(2/\pi)\cos^{-1}(-\zeta/u)$, and its image under
$\alpha\to-\alpha$ merge together to become a single
minimum at $\alpha=0$.
Thus we can conclude that the
phase transition across the $\zeta<0$, $u=|\zeta|$ line is second order. As is
the characteristic of such a phase transition,
at $u=|\zeta|$
the first three $\alpha$ derivatives of $V(\alpha)$ vanish at
$\alpha=0$. The same result holds for 
the line $\zeta>0$, $u=|\zeta|$. On the other
hand, phase transition across the line $\zeta=0$, $u<0$ is first order, as the
location of the minimum of $V(\alpha)$ jumps discontinuously from $\alpha=0$ to
$\alpha=2$ across this line.

Note that for $u>|\zeta|$ the value of the potential at the minimum
$\alpha_{min}$ is given by:
\be \label{emone}
V(\alpha_{min})=-{1\over g} [{1\over 2} (R-R_c) +{\zeta^2\over
4(R-R_c)}]\, .
\ee
{}From this one can calculate the mass of the non-BPS D-particle as
follows. First of all we note that the total mass of the system for a
given value of $\alpha$ must be related to $V(\alpha)$ by an additive
constant:
\be \label{emtwo}
M(\alpha) = C + V(\alpha)\, .
\ee
$C$ is determined by demanding the $M(\alpha=0)$ reproduces correctly the
mass of the D-string $\bd$-string pair. Since using footnote \ref{ff1} 
we
see that $\pi\zeta$ corresponds to the shift of the NS 5-brane near $x=\pi
R$, the net distance of this NS 5-brane from the $x=0$  plane is
$\pi(R+\zeta)$. Thus the mass of the D-string $\bd$-string system,
obtained by multiplying their length by the tension, is given
by $(R+\zeta)/g$. This gives:
\be \label{emthree}
C = {1\over g}(R+\zeta) - V(\alpha=0) = {1\over 2g} (R+R_c)\, .
\ee
Thus the mass of the stable non-BPS D-particle for $R>R_c+{1\over
2}|\zeta|$ is given by:
\be \label{emfour}
M(\alpha_{min}) = C +V(\alpha_{min}) = {1\over g} (R_c - {\zeta^2\over
4(R-R_c)})\, .
\ee

\vfill\eject

\end{document}